\documentclass[final,5p,times,twocolumn]{elsarticle}

\usepackage{amssymb}
\usepackage{amsmath}
\usepackage{grffile}

\journal{Journal of Magnetism and Magnetic Materials}

\newcommand{\I}{{\rm i}}
\renewcommand{\le}{\leqslant}
\renewcommand{\ge}{\geqslant}
\newcommand{\kk}{\mathbf{k}}

\newcommand{\doubl}{\langle \! \langle}
\newcommand{\doubr}{\left\rangle \! \right\rangle}

\newcommand{\zu}{z_{\uparrow}}
\newcommand{\zd}{z_{\downarrow}}

\newcommand{\es}{e_{\rm s}(\kk)}
\newcommand{\esm}{e_{\rm s}(\kk_m)}
\newcommand{\ea}{e_{\rm a}(\kk)}
\newcommand{\easq}{e_{\rm a}^2(\kk)}

\begin{document}

\begin{frontmatter}



\title{Magnetic phase transitions and unusual antiferromagnetic states in the Hubbard model}


\author[Ekb,Urfu]{P.A. Igoshev}
\author[Izh]{M.A. Timirgazin}
\author[Izh]{A. K. Arzhnikov}
\author[Ekb]{V. Yu. Irkhin}

\address[Ekb]{Institute of Metal Physics, Russian Academy of Sciences, 620108 Ekaterinburg, Russia}
\address[Izh]{Physical-Technical Institute, 426000, Kirov str. 132, Izhevsk, Russia}
\address[Urfu] {Ural Federal University, 620002 Ekaterinburg, Russia}

\ead{valentin.irkhin@imp.uran.ru}

\begin{abstract}
Ground state magnetic phase diagrams of the square and simple cubic lattices are investigated for the narrow band Hubbard model within the slave-boson approach by Kotliar and Ruckenstein.
The transitions between saturated (half-metallic) and non-saturated   ferromagnetic phases as well as similar transition in antiferromagnetic (AFM) state are considered in the three-dimensional case.  Two types of saturated antiferromagnetic state with different concentration  dependences of sublattice magnetization are found in the two-dimensional case in the vicinity of half-filling: the state with a gap between AFM subbands and  AFM state with large electron mass. The latter state is hidden by the phase separation  in the finite-$U$ case.
\end{abstract}


\end{frontmatter}


\section{Introduction}\label{sec:intro}

As first demonstrated by Nagaoka, in the limit of
infinite Hubbard's repulsion $U$ the ground state for simple bipartite lattices in the nearest-neighbour approximation is a saturated ferromagnetic state for a low density $\delta $ of
current carriers (doubly occupied states (``doubles'') or empty  states (``holes'') in an almost half-filled band)~\cite{Nagaoka:1966}.
Nagaoka considered the stability of saturated ferromagnetic  state (sFM) and  found its spin-wave instability 
with increasing $\delta$ and decreasing $U$. Roth applied
a variational principle to this problem and obtained two critical
concentrations~\cite{Roth:1969}. The first one, $\delta _{\mathrm{c}}$, corresponds to
instability of saturated ferromagnetic state, and the second one, $\delta _{%
\mathrm{c}}^{\prime }$, to the second-order transition from non-saturated ferromagnetism into paramagnetic state.

Zarubin and Irkhin \cite{Zarubin:2004,Zarubin:2012}
have applied the $1/z$-expansion of the Green's functions
in the many-electron representation \cite{Hubbard-IV:1965,Irkhin:1994}
for the Hubbard model and
obtained an interpolation description of saturated and
non-saturated ferromagnetism.


When introducing the Heisenberg exchange $J$ ($t-J$ model) a tendency to antiferromagnetism occurs
since  the ground state at $n = 1$ is  AFM insulator.
The 
hole states in AFM matrix (for empty conduction band) in the nearest-neighbor hopping approximation  at $J = 0$ were found to be incoherent~
\cite{Brinkman:1970, Varma:1988, Kane:1989}.
For finite $J$ the states near the  band bottom form a narrow coherent band with small residue of order $|J/t| \ll 1$ and heavy mass $\sim |t/J|$ \cite{Kane:1989}.
 However, this picture is broken  by different ways: (i) in the presence of next-nearest neighbor hopping which strongly affects the form of magnetic order; (ii) for finite density of carriers
 which makes Neel AFM order to be unfavorable; (iii) for finite Hubbard $U$ when a large number of spin excitation can be involved.

The competition  of FM and AFM ordering results in occurrence of  spiral magnetic ordering \cite{Igoshev:2010} or the magnetic phase separation \cite{Visscher:1973, Igoshev:2010, Igoshev:2015}.
These results were obtained under the assumption that saturated ferromagnetism is the ground state at finite doping and sufficiently large $U$.
Here we present a more general physical picture taking into account finite next-nearest electron hopping which results, in particular,  in occurrence of an unusual correlated antiferromagnetic state even at infinite $U$.



\section{Formalism}\label{sec:formalism}
We consider the Hubbard model~\cite{Hubbard-I:1963}
\begin{equation}
      \label{eq:original_H}
      H=\sum_{ij\sigma\sigma'} t_{ij}\delta_{\sigma\sigma'} c^\dag_{i\sigma}c^{}_{j\sigma'}+U\sum_i n_{i\uparrow}n_{i\downarrow},
\end{equation}
with the electron hopping $t_{ij} = -t$ for the nearest neighbors and $t'$  for the
next-nearest neighbors (we assume $t>0$), $c^\dag_{i\sigma},c^{}_{i\sigma}$ are the electron creation and annihilation
operators, respectively, $n_{i\sigma}=c^\dag_{i\sigma}c^{}_{i\sigma}$, $i$ is the site number, $\sigma$ is the spin projection.

The local spin space rotation around $x$ axis, matching different site magnetization vectors along, say, $z$ axis,  by the angle $\mathbf{QR}_i$ (where $\bf
Q$ is a spiral wave vector, ${\bf R}_i$ is the site position) is applied for the consideration of plane magnetic spirals. This
maps the spiral magnetic state into an effective ferromagnetic one, but the hopping term in the Hamiltonian becomes
non-diagonal with respect to index $\sigma$: $t_{ij}\delta_{\sigma\sigma'}\rightarrow t^{\sigma\sigma'}_{ij}=\exp[\mathrm{i}\mathbf{Q}(\mathbf{R}_i-\mathbf{R}_j)\sigma^x]_{\sigma\sigma'}t_{ij}$ in Eq.
(\ref{eq:original_H}).
The Hartree--Fock treatment of the many--particle Coulomb interaction term replaces it to some effective field $U\langle n_{i\bar\sigma}\rangle $ which  mixes the averaged contributions from singly and doubly occupied states.
However, this is not satisfactory even qualitatively, especially at large $U$.

A simple way of taking into account the correlation
effects is an extension of the configuration space to a bosonic sector by introducing the {\it slave-boson}  annihilation (creation) operators $e_i(e_i^\dag)$, $p_{i\sigma}(p_{i\sigma}^\dag),
d_i(d_i^\dag)$ for empty, singly and doubly occupied states, respectively~\cite{Kotliar_SB:1986, Fresard:1992}.
The transitions between the site states originating from intersite electron transfer are now accompanied by corresponding transitions in bosonic sector.
The equivalence of the original and new description is achieved by
the replacement $c_{i\sigma}\rightarrow \mathsf{z}_{i\sigma}c_{i\sigma}$,
where $\mathsf{z}_{i\sigma} = (1 - d^\dag_i d^{}_i - p^\dag_{i\sigma}p^{}_{i\sigma})^{-1/2}\left(e^\dag_ip^{}_{i\sigma} + p^\dag_{i\bar\sigma}d^{}_{i} \right)(1 - e^\dag_i e^{}_i - p^\dag_{i\bar\sigma}p^{}_{i\bar\sigma} )^{-1/2}$, 
which extends the action of $c_{i\sigma}$ on the bosonic subspace language in conjunction with 
the constraints
\begin{equation}
      \label{eq:eta-constraint}
      e^\dag_ie^{}_i+\sum_\sigma p^\dag_{i\sigma}p^{}_{i\sigma}+d^\dag_id^{}_i=1,
\end{equation}
\begin{equation}
      \label{eq:lmb-constraint}
      d^\dag_id^{}_i+p^\dag_{i\sigma}p^{}_{i\sigma}=c^\dag_{i\sigma}c^{}_{i\sigma}.
\end{equation}
The presence of the constraints can be taken into account within the functional integral formalism via the Lagrange
multipliers (
$\eta_i$ for Eq.~(\ref{eq:eta-constraint}) and 
$\lambda_{i\sigma}$ for Eq.~(\ref{eq:lmb-constraint})) introduced into the action.

Within the saddle-point approximation $e_i,p_{i\sigma},d_i$ are replaced by $i$-independent $c$-numbers, and $\mathsf{z}_{i\sigma}$ by $z_\sigma =(d^2+p_\sigma^2)^{-1/2}(ep_\sigma+p_{\bar\sigma}d)(e^2+p_{\bar\sigma}^2)^{-1/2}\le 1$.
Then the thermodynamical potential $\Omega$ 
has the form
\begin{equation}\label{eq:Omega_general}
	\Omega = \Omega_c + \Omega_b,
\end{equation}
where
	$\Omega_ c = -T\sum_{\nu\kk} \ln(1 + \exp(-\beta (E_\nu(\kk)-\mu)))/N$,
	$\Omega_b =  - 2\lambda d^2 - \lambda (p^2_{\uparrow} + p^2_\downarrow) + \Delta (p^2_{\uparrow} - p^2_\downarrow)$, $\mu$ being chemical potential and
\begin{equation}
	E_{\nu}(\kk)=(z^2_\uparrow+z^2_\downarrow)\es/2+\lambda + (-1)^\nu\sqrt{D_\kk},
\end{equation}
are eigenvalues of effective fermionic Hamiltonian
\begin{equation}\label{eq:H_eff}
	H^c_{\sigma\sigma'}(\kk) = \lambda - \Delta\sigma^z_{\sigma\sigma'} + z_\sigma z_{\sigma'}(\es\delta_{\sigma\sigma'} + \ea\sigma^x_{\sigma\sigma'}),
\end{equation}
\begin{equation}\label{eq:Dk_def}
D_\kk=\left((z^2_\uparrow-z^2_\downarrow)\es/2 - \Delta \right)^2+(\ea z_\uparrow z_\downarrow)^2
\end{equation}
and
    $e_{\rm s,a}(\kk) =(t_{\kk+\mathbf{Q}/2}\pm t_{\kk-\mathbf{Q}/2})/2$.
For convenience we have introduced
$\lambda=(\lambda_\uparrow+\lambda_\downarrow)/2$,
$\Delta=-(\lambda_\uparrow-\lambda_\downarrow)/2$.
Direct calculation of the action extremum with respect to boson variables and Lagrange multipliers yields SBA equations, see \cite{Igoshev:2015}.
We introduce the electronic density $n \equiv \sum_\sigma\langle n_{i\sigma}\rangle = \sum_\sigma p^2_\sigma + 2d^2$, and amplitude of (sublattice) magnetization $m \equiv \sum_\sigma\sigma\langle n_{i\sigma}\rangle =  p^2_\uparrow - p^2_\downarrow$.

The electronic Green's function
\begin{equation}
	G_{\sigma\sigma'}(\mathbf{k}, E) =\frac1{N}\sum_{i} \exp(-\I\kk\mathbf{R}_{ij})\doubl \mathsf{z}_{i\sigma}c_{i\sigma} |\mathsf{z}^\dag_{j\sigma'}c^\dag_{j\sigma'} \doubr_E
\end{equation}
is replaced  in  spirit of SBA by
\begin{equation}
	G_{\sigma\sigma'}(\mathbf{k}, E) = z_\sigma z_{\sigma'}
	\left(E - H^c_{\sigma\sigma'}(\kk)\right)^{-1}_{\sigma\sigma'} + G^{\rm inc}_{\sigma\sigma'}(\mathbf{k}, E),
\end{equation}
where $G^{\rm inc}$ contains both the incoherent contributions to the Green's function and the contribution of the interaction of electrons with well-defined collective excitations~\cite{Kane:1989}.
Using the Bogolubov transformation which diagonalizes the $H^c$ (see Eq.~(\ref{eq:H_eff})) 
	$c_{\kk\sigma} = \sum_{\nu}T_{\sigma,\nu}(\kk)\alpha_{\kk\nu}$,
we obtain the expression for contribution of coherent part of the Green's function 
\begin{equation}
\sum_{\sigma\sigma'}G^{\rm coh}_{\sigma\sigma'}(\mathbf{k}, E) = \sum_\nu \frac{a_\nu}{E - E_\nu(\kk)},
\end{equation}
\begin{equation}\label{eq:local_z}
 	a_\nu(\kk) = \sum\nolimits_{\sigma\sigma'}z_{\sigma}z_{\sigma'}\bar{T}_{\sigma,\nu}(\kk)T_{\sigma',\nu}(\kk)
\end{equation}
are bilinears  of {\it local} residues $z_\sigma$.
The loss of the quasiparticle weight in coherent states is seen from the sum rule
\begin{equation}\label{eq:sum_rule}
	\int \rho(E)\, dE = \sum\nolimits_{\nu\kk}a_\nu(\kk) = \sum\nolimits_{\sigma}z_\sigma^2,
\end{equation}
where $\rho(E) = -\pi^{-1}\sum_{\kk\sigma\sigma'}{\rm Im}G^{\rm coh}_{\sigma\sigma'}(\kk, E)$ is the coherent contribution to the   density of states~(DOS).

For infinite $U$ and $n < 1$ we have $d = 0$, so that $e^2 = \delta = 1 - n$; for $n > 1$ we have  $e = 0$ and should put $d^2 = \delta = n-1$. However, for simplicity we present the formulas in terms of $e^2$ only (note that the results for $n > 1$  are obtained from those for $n <1$ by the replacement $n\rightarrow 2-n, t'\rightarrow - t'$).
Unlike HFA approximation ($\Delta_{\rm HFA} = Um/2$), the solution for $\Delta$ becomes bounded even at $U \rightarrow\infty$.

Now we consider analytically the important case of AFM (or spiral) order at small number of holes
($\delta \rightarrow 0$).
Since $(\zu^2 - \zd^2)\es/2 - \Delta < 0$ for  most $\kk$-points in the Brillouin zone we can expand  (\ref{eq:Dk_def}) in $\zu\zd$.
Behavior of $p_\downarrow$
 depends dramatically on the value of the lattice sum
\begin{equation}
	C = \frac1{N}\sum_{\kk}\frac{\easq}{(e_{\rm s}(\kk_m) - \es)^2},
\end{equation}
$\kk_m$ being the position of maximum of lower ($\nu = 1$) subband.


For $\mathbf{Q} = \mathbf{Q}_{\rm AFM}$ ($\mathbf{Q}_{\rm AFM}= (\pi,\pi)$ for the square and $\mathbf{Q}_{\rm AFM}= (\pi, \pi, \pi)$ for the simple cubic (sc) lattice) we have
 $\es\propto t'$,  and $C$ decreases as $|t'|$ increases.
At $C < 1$  we get $p_\downarrow \propto e$ and
\begin{equation}\label{eq:C_expr}
\omega\equiv \lim_{e\rightarrow 0} (p^2_\downarrow/e^2) = C/(1 - C).
\end{equation}
At the same time, $\zu^2 \rightarrow (1 + \omega)^{-1}$ is finite, and $\zd^2 = e^2$ in the limit $\delta \rightarrow 0$,
so that direct AFM gap
	$\Delta = \esm(1 - C)/2$,
does not vanish.
%


In the  case $C > 1$ the equation (\ref{eq:C_expr}) is violated 
and actually $p_\downarrow \propto \sqrt{e}$.
In this case $\lim_{e\rightarrow 0}(p^4_\downarrow/e^2) = x^2$, where 
\begin{equation}\label{eq:x}
	x^2 = \frac{\kappa + e_s(\kk_m)}{N^{-1}\sum_{\kk}\easq(\es + \kappa)^{-2}[e_{\rm a}^2(\kk)/(\es + \kappa) - \es]},
\end{equation}
and $\kappa$ satisfies
\begin{equation}\label{eq:delta_equation}
	(1/N)\sum_{\kk}\easq/(\kappa + \es)^2 = 1.
\end{equation}
This results in $1 - m \sim  2x\sqrt{\delta}$, and both $\Delta \sim -\kappa\sqrt{\delta/x}/2$ and $\zu^2 \sim \sqrt{\delta/x}$ vanish as  $\sqrt{\delta}$. 

Direct calculation of the lattice sum $C$ allows to determine the character of  AFM state in the vicinity of half-filling.
For the square lattice with  $t'<0$ we find  $C > 1$   at  $0 < -t' < -t_{\rm c} = t/\sqrt{2\pi}\approx 0.4t$  and $C < 1$ otherwise. For $t' > 0$, $e_{\rm a}(\kk_m)\ne 0$, and $C$ always diverges  which is connected with the stability of sFM state.
For the sc lattice we have  $C < 1$ which implies ``usual'' antiferromagnetic behavior.

To consider the competition of sFM and AFM state in the limit $\delta\rightarrow0$ we expand the free energy $\mathcal{F} = \Omega + \mu n$ of spiral state (see Eq.~(\ref{eq:Omega_general})) by $\delta$
\begin{equation}\label{eq:F_expansion}
	\mathcal{F}_{\rm AFM} = \delta \left(\kappa + (1/{N})\sum\nolimits_{\kk} {\easq}/{(\kappa + \es) } \right) + o(\delta),
\end{equation}
where $\kappa$ is a solution to the equation (\ref{eq:delta_equation}) in the case $C > 1$ and $-\esm$ otherwise.
For sFM (Nagaoka) state
\begin{equation}\label{eq:F_Nagaoka_expansion}
	\mathcal{F}_{\rm sFM} = -4\delta(t + t') + o(\delta).
\end{equation}
The expansion of Eq.~(\ref{eq:F_expansion}) up to $\delta^2$  yields a description of phase separation  (PS) into (almost) uncorrelated AFM state at $\delta = 0$ and strongly correlated AFM state at finite doping.

\section{Results}

\begin{figure}[!h]
  \center
\includegraphics[width=0.49\textwidth]{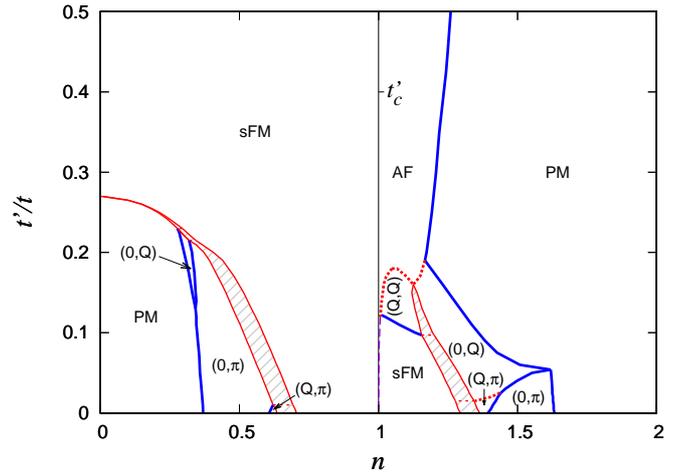}
\caption{
	(Color online)
	Ground state magnetic phase diagram of the Hubbard model with infinite $U/t$ for square lattice within SBA.
	The spiral phase regions are denoted according to the form of their wave vector (concrete {\it number} $Q$ depends on the point $(n,t')$ of the region).
	Filling shows the phase separation regions.
	Bold (blue) lines denote the second-order phase transitions.
	Solid (red) lines correspond to the boundaries between the regions of the homogeneous phase and phase
separation.
	Bold dashed (red) lines denote the first order phase transitions in the case where the region of the phase separation is narrow.
	Dashed (red) horizontal lines separate the phase separation regions corresponding to different phase pairs.
	The boundary  of AFM phase at very  small $n-1>0$ which is stable with respect to sFM phase (see  discussion at the end of Section \ref{sec:formalism} is shown schematically  by long-dashed (violet) line since this boundary is not found numerically because of precision problems at extremely small carrier density.
}
\label{fig:phase_diagram_sq_inf}
\end{figure}

We start from the case $U=\infty$ making the focus on the properties of the system in the close vicinity of $n = 1$ (small $\delta$).
The square lattice ground state diagram, calculated via the comparing $\Omega$ for different phases  in terms of $n$ and $t'\ge 0$ is depicted in Fig.~\ref{fig:phase_diagram_sq_inf}. It contains the regions of all the commensurate magnetic phases (antiferromagnetic (AFM) and ferromagnetic (FM)), the spiral magnetic states and the paramagnetic phase as well.

For $t'=0$ the picture is symmetric with respect to the half-filling due to the particle-hole symmetry.
Antiferromagnetic state at $n=1$ is replaced by saturated FM state at arbitrarily small doping in accordance with Nagaoka theorem~\cite {Nagaoka:1966}. At moderate doping $\delta\sim0.3$ FM phase goes to the spiral $(Q,\pi)$ structure through the first order transition with PS. $(Q,\pi)$ state smoothly transforms to the $(0,\pi)$ order, which corresponds to a layered antiferromagnet. At large doping $\delta\sim0.6$ the magnetic order becomes suppressed and the second order transition to the paramagnetic (PM) phase occurs.

Finite $t'$ values destroy the particle-hole symmetry and the diagram becomes strongly asymmetric. In the hole-doped half of the diagram the FM phase gradually displaces other states with increase of $t'/t$ and at $t'\gtrsim0.27t$ it occupies all the $n<1$ region. For the electron-doped half of the diagram ($n>1$) the FM phase region, on the contrary, becomes narrower with increase of $t'/t$ eventually being replaced by the diagonal spiral $(Q,Q)$ order phase region at $t'\sim (0.1-0.15)t$.
The regions of spiral magnetic phases, adjoining to ferromagnetic regions through the phase separation, are narrowed similarly with FM regions  up  to $t'\sim0.17t$ eventually being replaced by AFM state. Further increase of $t'$ yields the boundary of AF and PM states being weakly dependent on $\delta$. 

Generally, the instabilities of sFM state are (i) instability with respect to collective magnetic (spiral or AFM) excitations (typically accompanied by first order phase transition); (ii) spin-flip instability resulting in the unsaturated FM state formation. 
Within the SBA the necessary stability condition of sFM with respect to the second type of instability	is $\varepsilon_2 \equiv  \min\limits_{\kk}E_2(\kk) < \mu$, $\varepsilon_2$ being the bottom of upper subband. We define also $\epsilon_1 = \max\limits_{\kk}E_1(\kk)$ which is the top of lower  subband.
Since saturated ferromagnetic states implies vanishing of both $p_\downarrow$ and $d$, the expansion of SBA equations yields
$d/p_\downarrow = u + \sqrt{1 + u^2}$,
with
	$u = (1/2)\left({U}(E_{\rm kin}/(ep_\uparrow) + \varepsilon_{\rm B}(ep_\uparrow))^{-1} - e/p_\uparrow + p_\uparrow/e\right)$,
where $\varepsilon_{\rm B}$ is the bottom of bare band and $E_{\rm kin} = \sum_{\kk}t_{\kk}f_{\kk}$ is kinetic energy of electrons in the spin up subband. In this case quasiparticle residues coincides with local ones: $z^2_\uparrow = 1$, $z^2_\downarrow = \delta(1 + u p_\uparrow/e)^2/(1 + u^2)$.

The instability of sFM (Nagaoka)
state with respect to uFM can occur far from nesting features of electronic spectrum and van Hove singularities of bare DOS, favoring  saturated ferromagnetism, and is actually absent in Fig.~\ref{fig:phase_diagram_sq_inf}.

\begin{figure}[!h]
\includegraphics[width=0.5\textwidth]{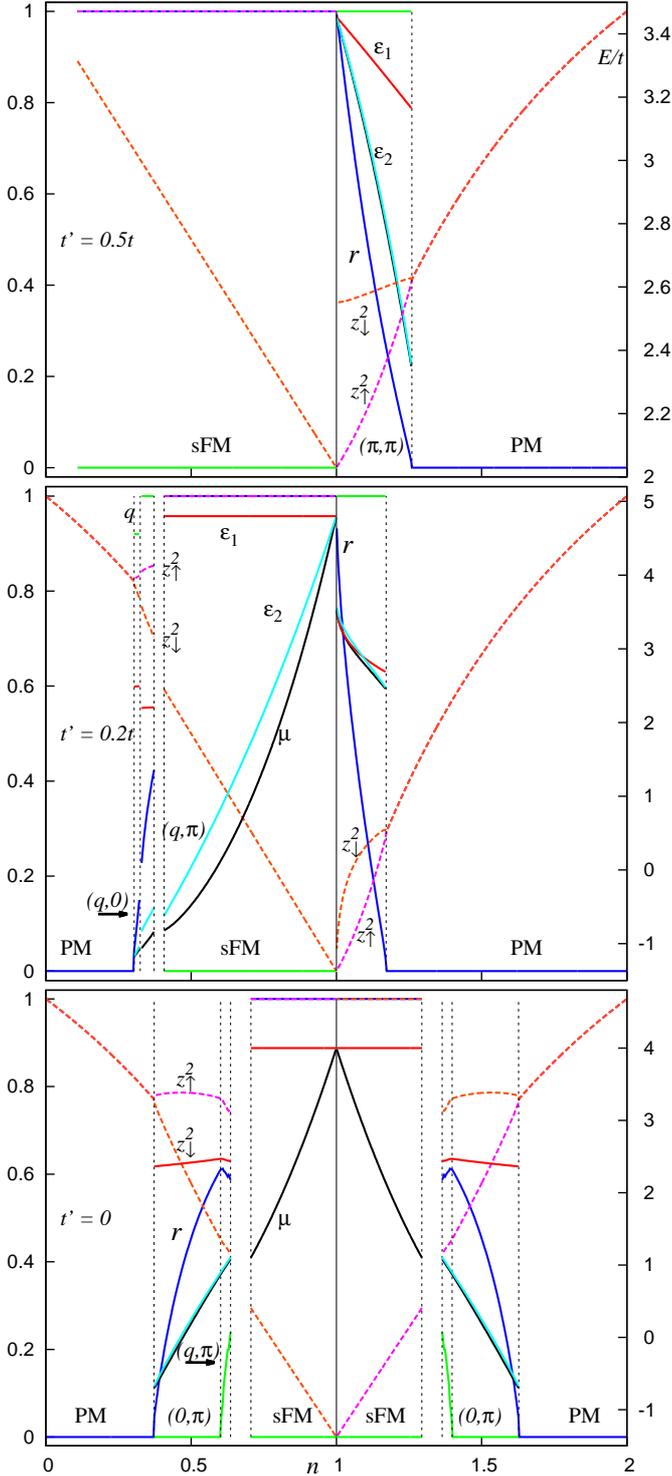}
\caption{
	(Color online)
    Left axis: the density	dependence of $q = Q_x$ (green), $\mu$ (black), relative magnetization  $r = m/(1-|1-n|)$ (blue),  $\zu^2$ (dashed violet), $\zd^2$ (dashed orange line). Right axis: $\varepsilon_1$ (red), $\varepsilon_2$ (light blue line). Lower panel presents $t' = 0$, middle panel $t' = 0.2t$, upper panel $t' = 0.5t$. Phases are denoted at the  panel bottom. Vertical dashed lines denote the region of discontinuous  phase transition (phase separation).
}
\label{fig:quantities}
\end{figure}

The different quantities $n$ scans  at fixed $t'=0, 0.2t, 0.5t$ corresponding to Fig. \ref{fig:phase_diagram_sq_inf} are presented in Fig.~\ref{fig:quantities}.
In the PM region, far away from half-filling, the local residues $z^2_\sigma$ demonstrate a  typical square-root dependence like that in the Brinkman-Rice theory of metal-insulator transition~\cite{Brinkman_MI:1970,Kotliar_SB:1986}.

The relative magnetization $m/(1-|1-n|)$ at $n < 1$ appears to be bounded from above by about 0.6 for the spiral state ($\mathbf{Q} = (0,\pi)$ or $\mathbf{Q} = (q,\pi)$)   which forms well away of half-filling.
These results are strongly different from  HFA results where large AFM gap $2\Delta_{\rm HFA} = Um$ causes $m/(1-|1-n|)\sim 1$.
The spiral states exist only in a small density interval being unstable with respect to sFM state when density becomes closer to $n = 1$.

The picture is strongly different in the case $n > 1$. The point of instability with respect to spiral (AFM) state of paramagnetic phase shifts towards to $n = 1$ with increasing $t'$.
The phase region of AFM state is rather narrow, whereas the sFM region is fully absent.
Generally, the correlated AFM phase possesses small $z^2_\sigma$ which is related to the transfer of the most of spectral weight into incoherent states. 
While $z^2_\downarrow$ linearly tends to zero as $\delta\rightarrow 0$, $z^2_\uparrow$ behaves differently depending on the value of $t'$: for $t'=0.2t$ $z^2_\uparrow \sim \sqrt{\delta}$, whereas for $t' = 0.5t$ $z^2_\uparrow$ tends to finite value.
We stress the difference of the behavior of AFM gap $2\Delta$ in the limit $\delta\rightarrow 0$  for $t' = 0.2t$ and $t' = 0.5t$.
While at $n \ne 1$ typically $\varepsilon_1 > \epsilon_2$ (the absence of the gap between the subbands), we find that in the case $t' = 0.2$ in the close vicinity of half-filling ($\delta<0.09$) the AFM state has a gap between subbands.
Another interesting consequence of this difference is the different asymptotics for sublattice magnetization:  $1 - m\propto \sqrt{\delta}$ for $t' = 0.2$ and $1 - m\propto \delta$ which agrees with the above analytics.


The vanishing of spectral weight in the system with small  $|t'|$ agrees with the results of earlier investigations of the motion of hole in AFM matrix~\cite{Varma:1988, Kane:1989} within the $t-J$ model in the nearest-neighbour approximation. They found that for $J=0$ the spectrum is incoherent, and  for finite $J$   a narrow coherent peak with small residue of order $J/t \ll 1$
occurs near the  band bottom.
Introducing  small ``direct'' exchange $J$ (e.g.~via the superexchange mechanism),
 yields a cutoff of divergence in Eq.~(\ref{eq:C_expr}), so that $\Delta\rightarrow \Delta + J m/2$ and $z^2_\uparrow$ becomes finite near half-filling. A similar cutoff takes place in the finite $U$ Hubbard model where effectively $J\sim t^2/U$.

\begin{figure}[!h]
  \center
\includegraphics[angle=-90,width=0.49\textwidth]{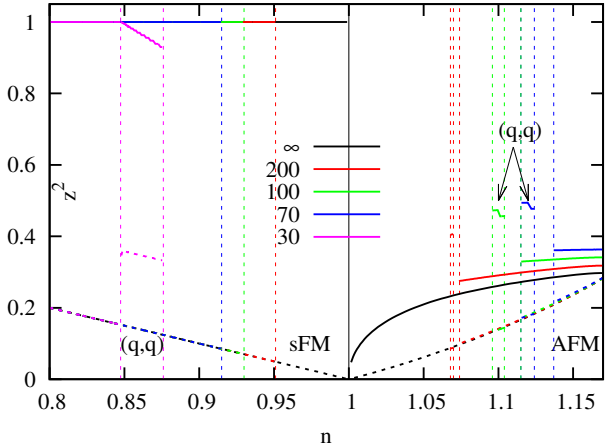}
\caption{
	(Color online)
		$z^2_\uparrow$ (solid line) and $z^2_\downarrow$ (dashed line) factors for the square lattice at $U/t = 30, 70, 100, 200, \infty$, $t' = 0.2t$.	Except for the $(q,q)$ phase, we have at $n<1$ sFM phase, and at $n>1$ AFM phase.
The breaks at vertical dashed line corresponds to boundaries of PS regions.
}
\label{fig:z_at_t1=0.2}
\end{figure}

\begin{figure}[!h]
  \center
\includegraphics[angle=-90,width=0.49\textwidth]{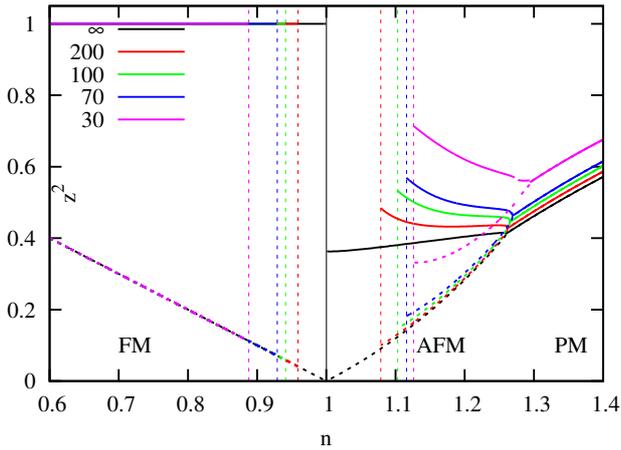}
\caption{
	(Color online)
		The same as in Fig.~\ref{fig:z_at_t1=0.2} for $t'=0.5t$.	
}
\label{fig:z_at_t1=0.5}
\end{figure}

Now we consider in detail the influence of finite values of $U$ on the properties of the system.
\begin{figure}[!h]
  \center
\includegraphics[width=0.49\textwidth]{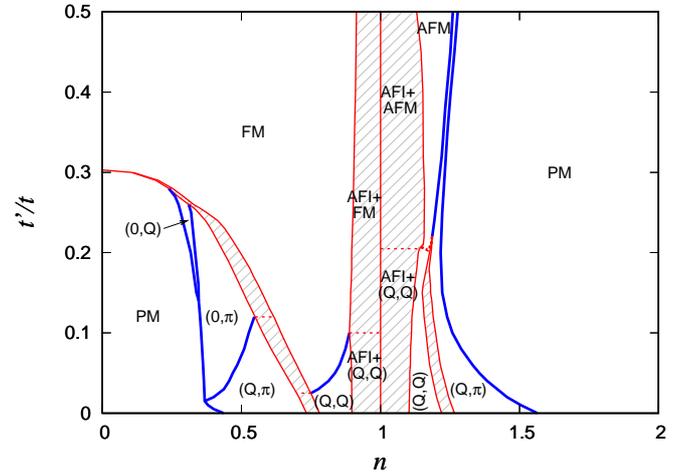}
\caption{
	(Color online)
		The same as in Fig.~\ref{fig:phase_diagram_sq_inf} for the square lattice at $U = 50t$.	 AFI is antiferromagnetic insulator state at $n = 1$.
}
\label{fig:phase_diagram_U50}
\end{figure}
In Fig.~\ref{fig:phase_diagram_U50} the ground state magnetic phase diagram at $U = 50t$ is presented. On can see that wide PS region occurs in the vicinity of $n = 1$. At $n < 1$ we find the PS into HFA-like AFM insulator and sFM state which width satisfies earlier estimate~\cite{Igoshev:2010, Visscher:1973}
\begin{equation}\label{eq:Visscher}
	\delta < \delta_{\rm PS} = \sqrt{{2t}/{[\pi (1+2t'/t) U]}},
\end{equation}
At the same time, at $n > 1$  
sFM state becomes unstable in the vicinity of half-filling with respect to the formation of 
AFM state with partially suppressed quasiparticle weight: $\zd^2\sim \delta$ or $\sqrt{\delta}$.
Thus AFM (or spiral) state occurs at arbitrarily large $U$ at $t' < 0$ which strongly changes the results by hiding the region of non-Fermi-behavior (with the size estimated as $\delta\sim ~ t/U$\cite{Kotliar:1988}).
To consider in detail the properties of the states taking part in PS we present the density dependence of 
$z$-factors (Figs.~\ref{fig:z_at_t1=0.2} and~\ref{fig:z_at_t1=0.5}) for different $U$.

We find that at $n<1$ $\Delta$ is almost insensitive  to $U$ and is nearly the same in both  sFM and spiral phases; the position of instability of PM state ($t'=0.2$) with respect to spiral phase is also almost fixed. In the sFM state $\Delta$ is much larger than its AFM value at $n>1$ ($t'<0$) which decreases with increasing  $U$ for both $t'=0.2t$ and $0.5t$. In the close vicinity of half-filling we obtain  quite   different behavior: at $t'= 0.2$ a precursor of unusual AFM behavior is found, $\Delta$ tends to saturation as $\delta$ arrive at $0$; at $t'=0.5$ it increases almost linearly. Both these dependences take a place until  PS occurs. While $z_\downarrow^2\sim \delta$ irrespective of $U$,  $z_\uparrow^2$ behavior always  depends strongly on $t'$: at $t' = 0.2t$ we find a decrease of $\zu^2$ with $\delta$ which is guessed as a  precursor of square-root vanishing at $U = \infty$ (unusual AFM behavior,  hidden  by PS). Note that the gap between AFM subbands exists  ($\varepsilon_1 < \varepsilon_2$)  in some $\delta$ region at large enough $U<\infty$. At $t' = 0.5t$ we find only a weak decrease of $\zu^2$ with increasing $U$, the dependence on $\delta$ being also weak.


The instability  of sFM with respect to the bound state of hole and spin flip  on the square lattice  was considered in \cite{Oles} where the energies of the states were compared in the framework of a  variational principle. It was found that sFM phase become unstable at $t' < -0.255t$. This conclusion was supported by DMRG study \cite{DMRG} where sFM phase was found to be stable up to $t'> -0.214t$ \textit{and} $n < 0.99$.
We  see that these DMRG results, although are reproduced at $\delta \gtrsim 0.01$ should be reconsidered  at smaller hole concentrations.
Direct calculation of free energies of sFM and AFM  state in limit $\delta\rightarrow0$ for the square lattice  using Eqs.~(\ref{eq:F_expansion}) and (\ref{eq:F_Nagaoka_expansion}) indicates favourability  of AFM state (this state is not shown in Fig.~\ref{fig:phase_diagram_sq_inf} due to precision problems at very small $\delta$).

\begin{figure}[!h]
\includegraphics[width=0.5\textwidth]{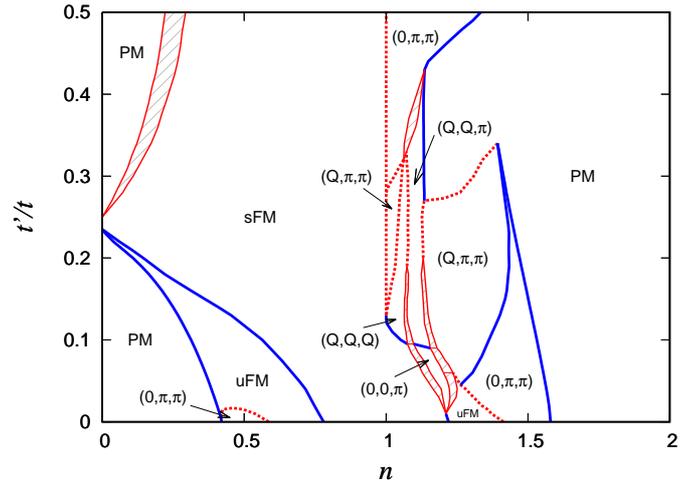}
\caption{
	(Color online)
	The phase diagram for sc lattice, the notations being as in Fig.~\ref{fig:phase_diagram_sq_inf}; uFM denotes the region of non-saturated ferromagnetic state.
	$\mathbf{Q}_{\rm AFM} = (\pi,\pi,\pi)$.
}
\label{fig:phase_diagram_sc_inf}
\end{figure}

\begin{figure}[!h]
\includegraphics[angle=-90, width=0.5\textwidth]{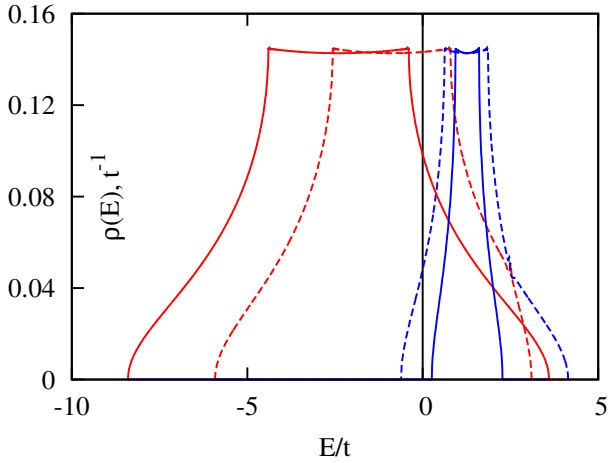}
\caption{
	(Color online)
	Spin resolved DOS in the vicinity of the sFM--uFM transition for sc lattice, $t' = 0$ for $U = \infty$ from sFM (solid lines, $n = m = 0.83, \zu^2 = 1, \zd^2 = 0.17$) and uFM (dashed lines, $n = 0.70, m = 0.59, \zu^2 = 0.84, \zd^2 = 0.31$). Spin up (down) contributions are shown by red (blue) lines.
}
\label{fig:spin_DOS}
\end{figure}

The phase diagram for the simple cubic lattice is shown in Fig.~\ref{fig:phase_diagram_sc_inf}.
Whereas for the square lattice  ferromagnetism is always saturated owing to influence of the logarithmic van Hove singularity, the magnetic phase diagram of the sc lattice contains the region of unsaturated ferromagnetism.
 An  example of spin-resolved density of states in the vicinity of transition from saturated ferromagnetic (``half-metallic'', sFM) state to uFM state in shown in Fig.~\ref{fig:spin_DOS}.
 One can see that, besides band narrowing, a shift of spin subbands occurs~\cite{Harris-Lange:1967}, which favors occurrence of ferromagnetism, in contrast with the simple Hubbard-I approximation~\cite{Hubbard-I:1963}.
The behavior of spin-up states in the saturated ferromagnetic state coincides with that of free electrons, whereas spin-down states below the Fermi level are strongly incoherent~\cite{Irkhin:1985, Edwards:1970}. The latter states are disregarded in our approximation and are therefore absent in Fig. \ref{fig:spin_DOS}; they should be taken into account to restore above-discussed sum rule (\ref{eq:sum_rule}) for the density of states. It is remarkable that the amplitude of the peaks appears to be the same for both subbands.

\begin{figure}[!h]
\includegraphics[angle=-90, width=0.5\textwidth]{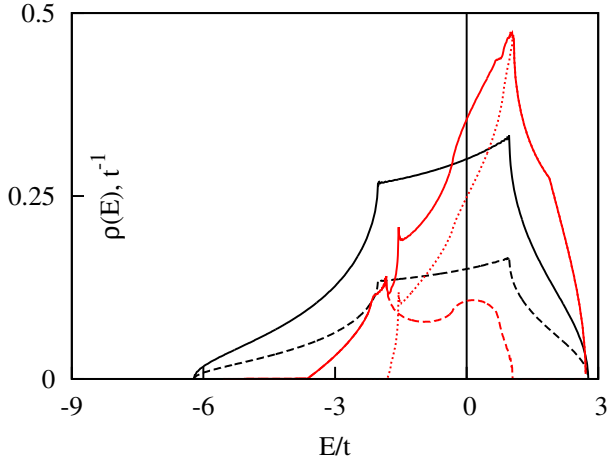}
\caption{
	(Color online)
	AFM subband resolved DOS for the sc lattice with $U = \infty, t' = -0.1t$ in the vicinity of transition from paramagnetic state~(black lines), $n = 0.4, z^2_\sigma = 0.75$ and  antiferromagnetic state~(red lines), $\mathbf{Q}_{\rm AFM} = (0, \pi,\pi)$, $n = 0.475, m = 0.068, \zu^2 = 0.72, \zd^2 = 0.66$. Solid line is total density of states, dashed (dotted) line is DOS for lower (upper) AFM subband. 	
}
\label{fig:AFM_transition1}
\end{figure}

\begin{figure}[!h]
\includegraphics[angle=-90, width=0.5\textwidth]{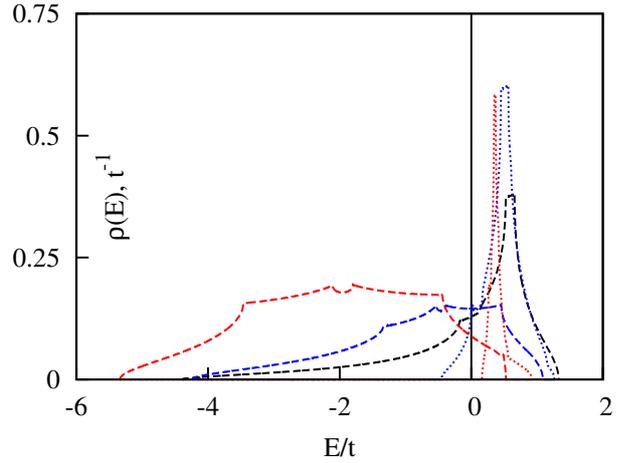}
\caption{
	(Color online)
	AFM subband resolved DOS for the sc lattice with $U = \infty, t' = -0.45t$  for paramagnetic phase (black line, $n = 0.78, z^2_\sigma = 0.38$), `non-saturated' AFM state  (blue line, $n = 0.8, m = 0.3, \zu^2 = 0.45, \zd^2 = 0.27$) and `saturated' AFM state (red line, $n = 0.94, m = 0.9, \zu^2 = 0.76, \zd^2 = 0.06$).
}
\label{fig:AFM_transition2}
\end{figure}

As discussed in Sect.~\ref{sec:formalism}, there is no heavy-electron AFM phases for sc lattice.
The behavior of $\varepsilon_1$ and $\varepsilon_2$ relatively to $\mu$ allows to introduce the classification of transition from saturated to non-saturated AFM state. The density of states
transitions driven by $\delta$ (from paramagnetic to non-saturated AFM and saturated AFM with $\mathbf{Q} = (0, \pi, \pi)$) for rather close points are shown in Fig.~\ref{fig:AFM_transition1} and \ref{fig:AFM_transition2}.
One can see that at small $|t'|$ (Fig.~\ref{fig:AFM_transition1}) upper and lower subbands  overlap considerably near the transition, whereas the energy dependence of density of states (DOS) strongly changes its form due to formation of the AFM order. For  large $|t'|=0.45$  another picture occurs: the transition from saturated to non-saturated AFM state results in broadening of upper subband and contraction of the lower one, which is caused by AFM order, similar to FM case. This similarity is a consequence of the fact at large $|t'|$ the electron transport includes to a large extent next-nearest neighbour sites with parallel spins. The main distinction with FM case is strong difference in amplitude of partial subband DOS's which is a consequence of $\kk$-dependent quasiparticle residue in AFM state. 


To conclude, we have presented the picture of magnetic phase transitions in the strongly correlated Hubbard model.
Although  HFA cannot yield reasonable results for the properties of the system at large $U/t$,  SBA results  provides a detailed information including considerable renormalization $z$-factors.
Further investigation with proper inclusion of the incoherent states and spin dynamics are required.

\section{Acknowledgments}
The research was carried out within the state assignment of FASO of Russia (theme ``Quantum'' No. 01201463332).
This work was supported in part by Ural Division of RAS (project no. 15-8-2-9, 15-8-2-12) and by the Russian Foundation for Basic Research (project no.
16-02-00995) and Act 211 Government of the Russian Federation 02.A03.21.0006.
The main amount of calculations was performed using the ``Uran'' cluster of IMM UB RAS.


\end{document}